\documentstyle[preprint,aps,prl,epsfig]{revtex}

\newcounter{listfig} 

\begin{document}


\title{Tunneling of magnetization versus spin--phonon and spin--spin
transitions in LiY$_{0.998}$Ho$_{0.002}$F$_{4}$}

\author{R. Giraud$^a$, W. Wernsdorfer$^a$, A.M. Tkachuk$^b$, D. Mailly$^c$,
and B. Barbara$^a$}

\address{$^a$ Laboratoire de Magn\'etisme Louis N\'eel, CNRS, BP166, 38042
Grenoble Cedex-09, France\\
$^b$ All-Russia Scientific Center ``S.I. Vavilov State Optical Institute'',
199034 St. Petersburg, Russia\\
$^c$ Laboratoire de Photonique et de Nanostructures, CNRS, 196 Av. H. Ravera, 
92220 Bagneux, France\\}

\date{\today}

\maketitle

\begin{abstract}

Strong hyperfine coupling in a $0.2\%$ Holmium doped LiYF$_{4}$ single 
crystal induces staircaselike hysteresis loops of the magnetization at very 
low temperatures. The field sweep rate dependence of hysteresis loops allows 
the study of two different regimes in the magnetic relaxation of these weakly 
coupled magnetic moments.
At slow field sweep rates, quantum tunneling of the magnetization occurs at 
avoided level crossings in the low-energy scheme of a single ion Ho$^{3+}$. 
At faster sweep rates, nonequilibrated spin--phonon and spin--spin 
transitions, mediated by weak dipolar interactions, lead to magnetization 
oscillations and additional steps. 


\end{abstract}


\narrowtext

Low frequencies quantum fluctuations of large magnetic moments, with a large 
uniaxial anisotropy, are reminiscent of mesoscopic magnetism. At very low 
temperature, the slow quantum dynamics associated with small tunneling gaps 
leads to staircaselike hysteresis loops. 
Discovered before in molecular magnets (see \cite{Igor} for a review), the 
quantum tunneling of the magnetization was recently observed in a weakly doped 
holmium fluoride, namely LiY$_{0.998}$Ho$_{0.002}$F$_{4}$, at slow field sweep 
rates and very low temperatures \cite{Giraud01}. As a result of a weak 
coupling to the cryostat at subkelvin temperatures, the sample is strongly 
sensitive to internal 
heating, due to phonon emission during the magnetization reversal.  An increase
of  the field sweep rate thus leads to a crossover from a quantum behavior at
low  rates to a magnetic relaxation regime dominated by spin--spin and
spin--phonon  interactions in a phonon bottleneck regime. In this paper, we
present field  sweep 
rate dependent hysteresis loops, at a cryostat temperature $T=40$~mK, which 
evidence this crossover. The weak dependence of hysteresis loops to a 
transverse applied field at fast sweep rates, compared to the strong 
dependence observed in the tunneling regime, also highlights this change in 
the nature of the dominant relaxation mechanism. 

\begin{figure}
\includegraphics[width=15cm]{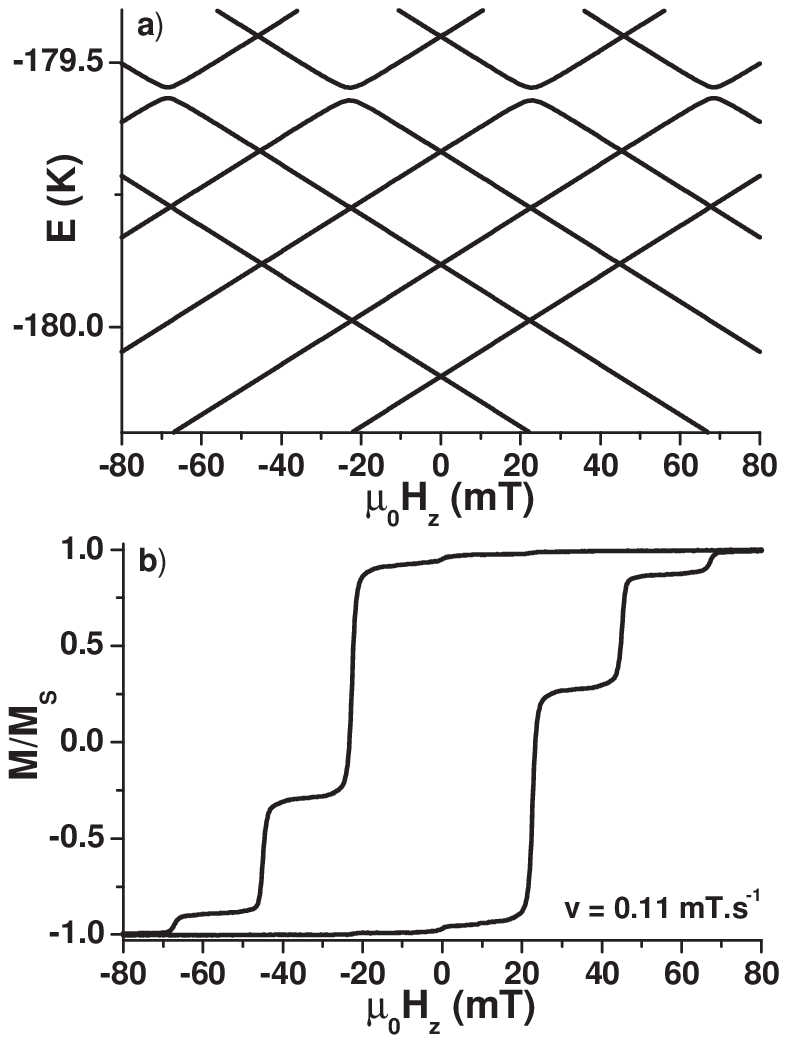}
\caption{R. Giraud}
\label{fig1}
\end{figure}

\begin{figure}
\includegraphics[width=15cm]{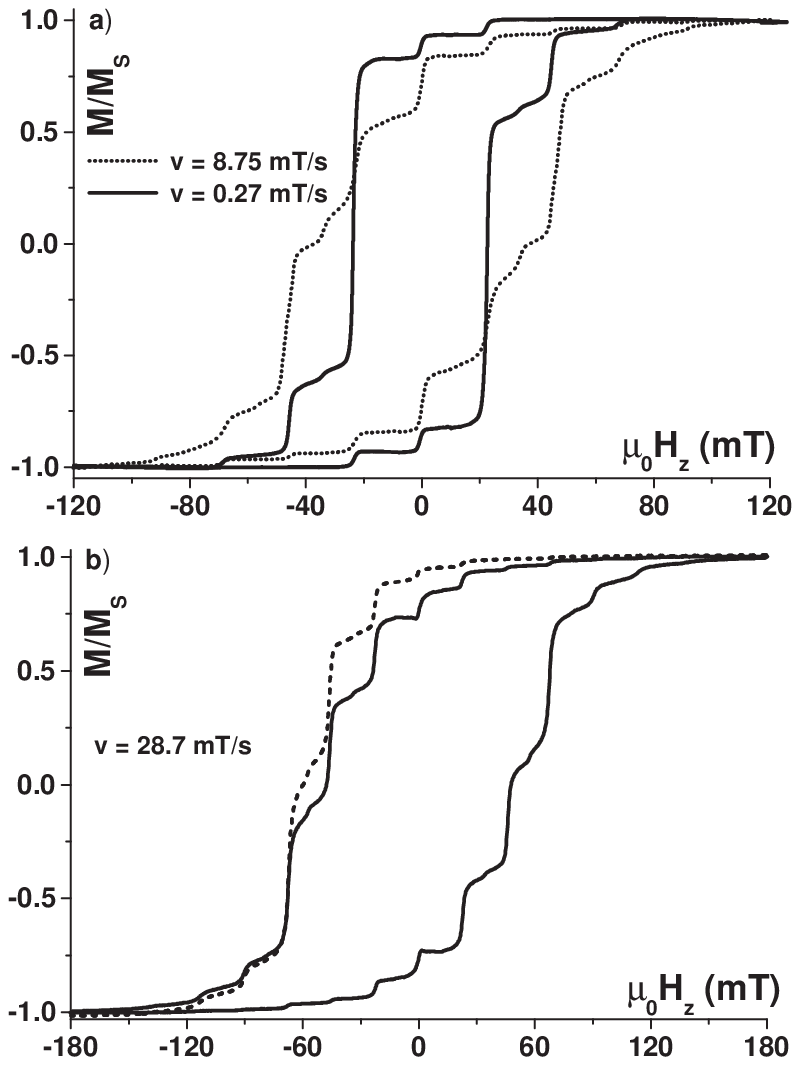}
\caption{R. Giraud}
\label{fig2}
\end{figure}

\begin{figure}
\includegraphics[width=15cm]{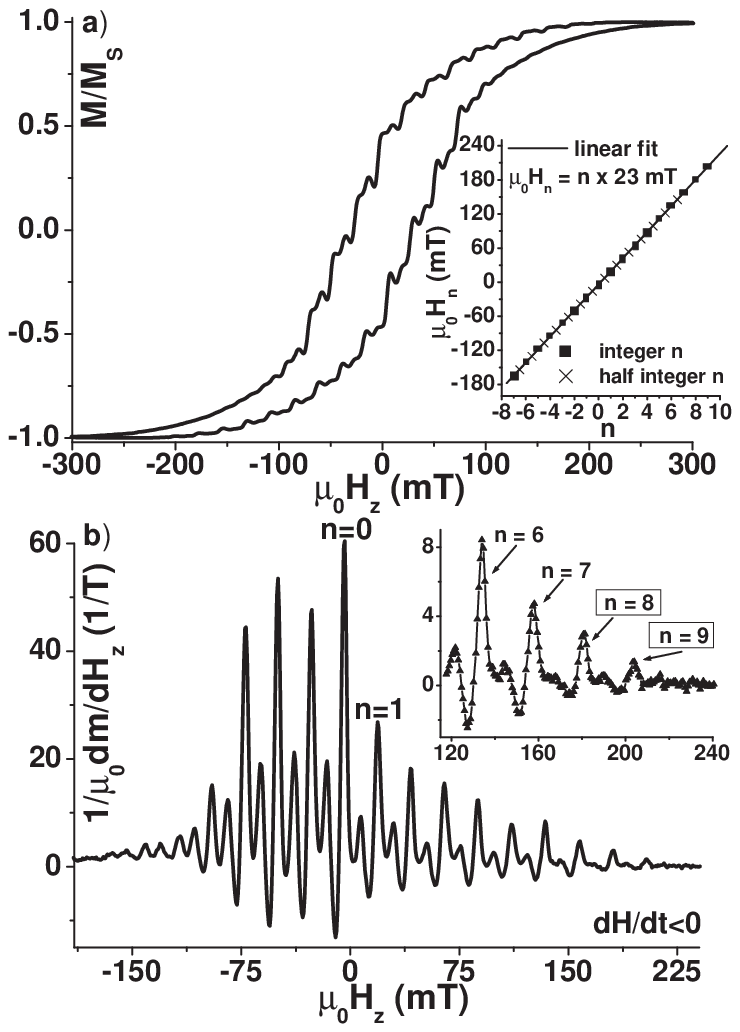}
\caption{R. Giraud}
\label{fig3}
\end{figure}

\begin{figure}
\includegraphics[width=15cm]{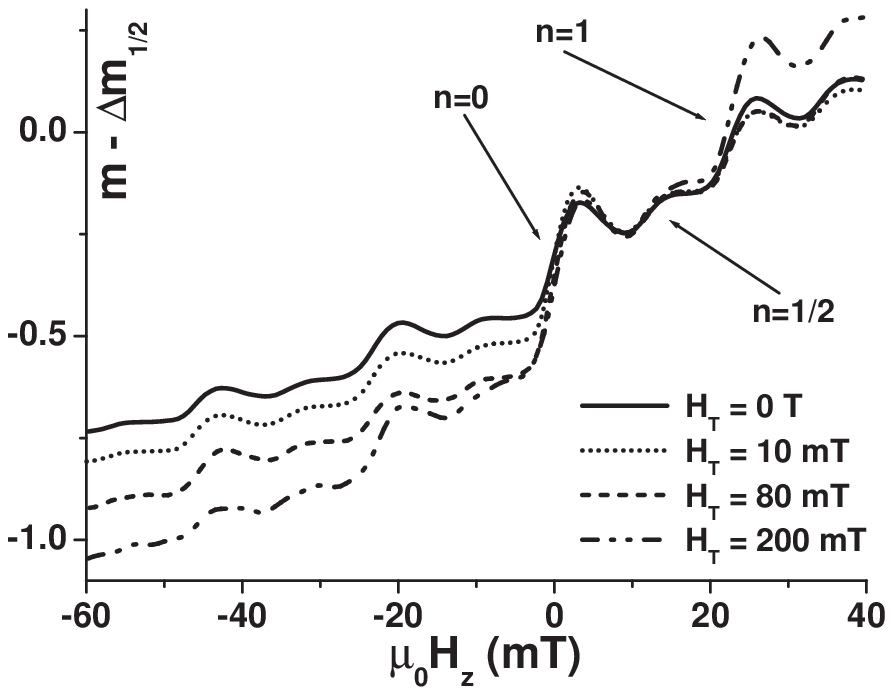}
\caption{R. Giraud}
\label{fig4}
\end{figure}

Diluted rare-earth ions in a nonmagnetic insulating single crystal are mainly 
investigated for applications in high-power laser diodes \cite{Tkachuk78}. 
In magnetism, very diluted samples are also relevant to study a nearly single 
ion quantum behavior of weakly coupled magnetic moments. The studied crystal 
has a tetragonal scheelite structure with S$_{4}$ point symmetry group at 
Ho$^{3+}$ sites. Due to the crystal field, each magnetic ion of $^{165}$Ho is 
characterized by an 
Ising-type ground state doublet. At very low temperatures, the system should 
be equivalent to a two-level system, with the magnetization lying along the 
$c$-axis. 
However, the electronic magnetic moment is strongly coupled to its nuclear 
spin ($I=7/2$) which split each ground state into eight levels, leading to 
level crossings in the Zeeman diagram for  resonant values
$\mu_0 H_n=n\times23$~mT ($-7\le n\le 7$), as already  discussed in
\cite{Giraud01}. The low excited states in this diagram are shown in
Fig.~\ref{fig1}a). Because of a crystal field induced hyperfine level repulsion,
there are some strongly avoided level crossings. Note that the degeneracies of
the other level crossings are also removed by internal fields fluctuations.   
      Magnetic measurements were made at $0.04<T<1$~K and for
$\mu_0H<2$~T, with a micro-SQUID magnetometer \cite{Werns97} allowing field
sweep rates up to 1~T/s. The crystal is first saturated in a large positive
field applied along the $c-$axis $\mu_0H_{\rm sat}\approx0.3$~T, and then the
field $H_z$ is swept between $\pm H_{\rm sat}$.  At slow field sweep rates and
very low  temperatures, quantum tunneling at avoided level crossings leads to 
staircaselike hysteresis loops, as shown in Fig.~\ref{fig1}b). The 
well-defined steps, observed at $T=40$~mK for a field sweep rate 
$v=0.11$~mT/s, agree with the energy level scheme shown above. This quantum 
relaxation is strongly enhanced  by a constant transverse field, as a result
of a rapid increase of the tunnel  splittings \cite{Giraud01}. Furthermore,
hysteresis loops are also  very sensitive to the field sweep rate, as shown in
Fig.~\ref{fig2}a). With  the increase of the sweep rate, magnetization steps
happen before field  inversion and hysteresis develop in larger applied
fields, showing the  influence of the phonon bath. In addition, the observed
hysteresis loops  depend on sample thermalization, showing the spin--phonon 
system is not at equilibrium with the cryostat (phonon bottleneck) 
\cite{AB,V15,Curtis}. 
When the field is swept back and forth after cooling the sample under 
$H_{\rm sat}$, a stationary regime occurs with reduced hysteresis compared to 
the first magnetization curve (Fig.~\ref{fig2}b). Such a behavior 
occurs 
because of phonon emission during the magnetization reversal. The phonon 
bottleneck induces an increase of the internal temperature in the sample, 
which depends on the sweep rate.
A hysteresis loop measured at $T=50$~mK for a much faster field sweep rate
($v=0.3$~T/s) is shown in Fig.~\ref{fig3}a). A succession of equally spaced
large and weak magnetization steps occur at fields $H_n$, with $-14\le2n\le14$.
The larger ones, with integer $n$, are associated with several equally spaced
level crossings and the smaller steps, with half integer $n$, fall just in
between when the levels are equally spaced. 
Equilibrium within the spin system is due to either quantum fluctuations at 
avoided
level crossings (integer $n$) or to spin--phonon transitions and/or 
cross-spin relaxation, allowed by weak dipolar interactions, when energy 
levels are almost equally spaced (integer and half integer $n$)
\cite{Curtis,Bloem59Hellwege68}. Spin--spin interactions allow two additional 
steps for $n=8$ and $n=9$, at fields with equally spaced levels but no level 
crossing (inset of Fig.~\ref{fig3}b). If the field sweep is suddenly 
stopped, 
the spin--phonon system exchanges energy with the cryostat and the 
magnetization relaxes toward the equilibrium curve.  
A small transverse applied field only increases the zero-field magnetization 
step (Fig.~\ref{fig4}), showing enhanced quantum fluctuations. Other resonances
and small magnetization  steps, dominated by cross-spin relaxation, are not
affected by a small  transverse field, if small enough ($\mu_0 H_T
\lesssim20$~mT). This is  emphasized in Fig.~\ref{fig4} where the small step
at $\mu_0 H_{1/2}=11.5$~mT  remains the same when increasing $H_T$, as do all
the magnetization steps  before field inversion if the transverse field is not
too large. As expected for the magnetization step at $\mu_0 H_1=23$~mT, which
involves a  large tunneling gap, the increase of this step requires larger
transverse  fields, as high as $200$~mT.

In conclusion, the field sweep rate dependence of hysteresis loops at very low 
temperatures clearly evidences a crossover between two different regime in the 
magnetic relaxation of weakly coupled magnetic moments in 
LiY$_{0.998}$Ho$_{0.002}$F$_{4}$. At slow sweep rates, quantum tunnelling of 
the magnetization occurs and is monitored by both crystal field and hyperfine couplings. 
At faster field sweep rates, additional magnetization steps are attributed to 
spin--phonon transitions and cross-spin relaxation in a phonon bottleneck 
regime. 

This work has been supported by DRET, Rh\^one-Alpes, MASSDOTS ESPRIT,
MolNanoMag TMR and AFIRST.

\newpage

\centerline{LIST OF FIGURES CAPTIONS}

\begin{list}{FIG.~\arabic{listfig}}{\usecounter{listfig}}

\item  Up a): Zeeman diagram of the split electronic ground state doublet by 
the hyperfine interaction (low-energy part). The level crossings occur for 
resonant values of the longitudinal field $H_n$. Down b): 
Hysteresis loops at $T\approx40$~mK and for $v=0.11$~mT/s showing quantum 
tunneling of the magnetization.

\item  Up a): Hysteresis loops for faster sweep rates at $T\approx40$~mK, 
$v=0.27$~mT/s (full line) and  $v=8.7$~mT/s (dotted line). Down b): Hysteresis 
loops at $T\approx40$~mK and for $v=28.7$~mT/s. As shown by the dashed line, a 
larger hysteresis occurs for the first magnetization reversal (smaller 
internal temperature).

\item  Up a): Hysteresis loops at $T=50$~mK and for $v=0.3$~T/s. Several
magnetization steps are observed for resonant values of the applied field
$\mu_0 H_n\approx n\times 23$~mT (see inset, $H_n$ values are deduced from
Fig.~\ref{fig3}b). Down b): Derivative of the loop shown in a) for a
decreasing field. The two additional measured steps shown in the inset, for
$n=8$ and $n=9$, are associated with cross-spin relaxation only.

\item  Part of the hysteresis loops at $T=50$~mK, for $v=0.3$~T/s and for 
several transverse applied fields (d$H_z$/dt>0). The shifted magnetization 
curves $m-\Delta 
m_{1/2}$, with $m=M/M_s$ and $\Delta m_{1/2}=m(H_{1/2},H_T)-m(H_{1/2},0)$, are 
used to highlight the small magnetization step at $\mu_0 H_{1/2}=11.5$~mT.  

\end{list}

\newpage

\newpage

\newpage

\newpage

\end{document}